\documentclass[12pt]{article}

\def\be{\begin{equation}}
\def\ee{\end{equation}}
\def\bea{\begin{eqnarray}}
\def\eea{\end{eqnarray}}
\def\ba{\begin{array}}
\def\ea{\end{array}}
\def\nn{\nonumber}

\evensidemargin \oddsidemargin

\begin{document}

\newcommand{\story}{\vspace{5mm} \noindent $\spadesuit$ }

%%%%%%%%%%%%%%%%%%%%%%%%%%% TITLE PAGE %%%%%%%%%%%%%%%%%%%%%%%%%%%%%%
\begin{titlepage}
%---------------------------- Title ---------------------------------
\vspace{15mm}
\baselineskip 9mm
\begin{center}
  {\Large \bf Fermions tunnelling from \\ the charged dilatonic black holes}
\end{center}

%--------------------- Authors and Addresses ------------------------
\baselineskip 6mm
\vspace{5mm}
\begin{center}
 De-You Chen\textsuperscript{a},
\quad Qing-Quan Jiang\textsuperscript{b,}\footnote
{E-mail:jiangqq@iopp.ccnu.edu.cn(Corresponding author)}
 and Xiao-Tao Zu\textsuperscript{a}\\
\textsuperscript{a}{\it School of Physical Electronics, University of Electronic Science}\\ {\it and Technology of China, Sichuan, Chengdu %
610054, China.}\\
\textsuperscript{b}{\it Institute of Particle Physics, Central China
Normal University,} \\{\it Wuhan, Hubei
430079, China.}\\
\end{center}

\thispagestyle{empty}
%\date{\today}

%-------------------------- abstract --------------------------------
\begin{center}
{\bf Abstract}
\end{center}
\noindent Kerner and Mann's recent work shows that, for an
uncharged and non-rotating black hole, its Hawking temperature can
be correctly derived by fermions tunnelling from its horizons. In
this paper, our main work is to improve the analysis to deal with
charged fermion tunnelling from the general dilatonic black holes,
specifically including the charged, spherically symmetric
dilatonic black hole, the rotating Einstein-Maxwell-Dilaton-Axion
(EMDA) black hole and the rotating Kaluza-Klein (KK) black hole.
As a result, the correct Hawking temperatures are well recovered
by charged fermions tunnelling from these black holes.
\\ [5mm]
\\[1mm]
\vspace{5mm}
\end{titlepage}

%%%%%%%%%%%%%%%%%%%%%%%%%%% BODY OF PAPER %%%%%%%%%%%%%%%%%%%%%%%%%%%
\baselineskip 7mm

%%%%%%%%%%%%%%%%%%%%%%%%%%%%%%%%%%%%%%%%%%%%%%%%%%%%%%%%%%%%%
\section{Introduction}
%%%%%%%%%%%%%%%%%%%%%%%%%%%%%%%%%%%%%%%%%%%%%%%%%%%%%%%%%%%%%

Since Hawking proved that a black hole can radiate thermally
\cite{SWH}, many papers have appeared to deeply discuss the
quantum radiation of black holes via different methods \cite{TGJ}.
Recently, Wilczek and his collaborators have proposed two
universal methods to correctly recover Hawking radiation of black
holes. In Ref.\cite{SF12}, when the classically irrelevant ingoing
modes at the event horizon of the black hole is neglected, the
effective chiral theory contains an anomaly with respect to
general coordinate symmetry, which is often named as gravitational
anomaly. To cancel gravitational anomaly and restore general
coordinate covariance at the quantum level, an energy momentum
tensor flux must be introduced at the horizon. The result shows
that the compensating energy momentum tensor flux has an
equivalent form as that of $(1 + 1)$-dimensional blackbody
radiation at the Hawking temperature. Later, much work further
promotes this method to the cases of different-type black
holes\cite{all1,all2,all3}.

On the other hand, Hawking radiation could be viewed as a
semi-classical quantum tunnelling process. According to the WKB
approximation, the tunnelling rate takes the form as
$\Gamma\propto \exp(-2\textrm{Im}I)$, where $I$ is the classical
action of the trajectory. Thus the calculation of the imaginary
part of the action becomes most important for this tunnelling
method. In general, two universal methods are applied in
references to derive the action. One method is called as the Null
Geodesic method(a detail description is available in
Ref.\cite{M}), and regarding the imaginary part of the action as
only contribution of the momentum $p_r$ of the emitted null
s-wave. Another method, proposed by Srinivasan and
Padmanabhan\cite{KS} and recently developed by Angheben
et.al\cite{M2}, is successfully present to derive the imaginary
part of the action by solving the Hamilton-Jacobi equation, which
is, later, called as the Hamilton-Jacobi method.

Till now, a lot of work has already been successfully carried out
for further development of the tunnelling approach, but all of
them are only focused on Hawking radiation of scalar particles
from various black hole spacetimes \cite{M3,M4}. In fact, a black
hole can radiate any types of particles at the Hawking
temperature, and the true emission spectrum should contain
contributions of particles with charge and all possible spins.
Recently, Kerner and Mann have succeeded to apply uncharged
fermion tunnelling from a non-rotating black hole to correctly
recover its Hawking temperature \cite{M5}. Subsequently, people
extend the analysis to the cases of Kerr black hole, Kerr-Newman
black hole and dynamical horizon and all the results are
satisfying\cite{M6}. In this paper, we further improve it to deal
with charged fermion tunnelling from the general dilatonic black
holes, specifically including the charged, spherically symmetric
dilatonic black hole, the rotating Einstein-Maxwell-Dilaton-Axion
(EDMA) black hole and the rotating Kaluza-Klein(KK) black hole.

The rest of the paper are organized as follows. In Sec.\ref{SSDB},
we study charged fermion tunnelling from the charged, spherically
symmetric dilatonic black hole, and the expected Hawking
temperature is well recovered. In Sec.\ref{EMDA} and Sec.\ref{KK},
for a broad extension, we again check charged fermions tunnelling
from the rotating Einstein-Maxwell-Dilaton-Axion (EMDA)and
Kaluza-Klein (KK) black holes. Sec.\ref{cd} contains some
discussions and conclusions.

%%%%%%%%%%%%%%%%%%%%%%%%%%%%%%%%%%%%%%%%%%%%%%%%%%%%%%%%%%%%%
\section{Fermions tunnelling from the charged, spherically symmetric dilatonic black hole}\label{SSDB}
%%%%%%%%%%%%%%%%%%%%%%%%%%%%%%%%%%%%%%%%%%%%%%%%%%%%%%%%%%%%%
In this section, we focus our attention on Hawking radiation of
fermions via tunnelling from the charged spherically symmetric
dilatonic black hole. Dilaton field is a kind of scalar field
which occurs in the low energy limit of the string theory where
the Einstein action is supplemented by fields such as axion, gauge
fields and dilaton coupling in a nontrivial way to the other
fields. Solutions for charged dilaton black holes in which the
dilaton is coupled to the Maxwell field have been obtained and
have many differences from that of ordinary black holes obtained
in the Einstein gravitational theory. Since the presences of
dilaton they have important consequences on the causal structure
and the thermodynamic properties of the black hole, and much
interest is attracted to study the dilaton black holes. The
spherically symmetric solution \cite{M7} is obtained from the
four-dimensional low energy Lagrangian
\begin{equation}
S= \int {dx^4\sqrt { - g} \left[ { - R + 2\left( {\nabla \Phi }
\right)^2 + e^{ - 2a\Phi }F^2} \right]} , \label{S2}
\end{equation}
where $a$ is a parameter, and denotes the strength of the coupling
of the dilation field $\Phi $ to Maxwell field $F$. When $a = 0$,
it reduces to the usual Einstein-Maxwell scalar theory. When
$a=1$, it is part of the low energy action of string theory. The
metric of the charged, spherically symmetric dilatonic black hole
(also called as G. H dilatonic black hole) reads as
\bea
&& ds^2 =
- e^{2U}\left( r \right)dt^2 + e^{ - 2U}\left( r \right)dr^2 +
R^2\left( r \right)\left( {d\theta ^2 + \sin ^2\theta d\varphi ^2}
\right),\nn\\
&& e^{2\Phi } = \left( {1 - \frac{r_ - }{r}} \right)^{\frac{2a}{1 +
a^2}}, \quad F = \frac{Q}{r^2}dt \wedge dr, \label{eqds} \eea where
$e^{2U}\left( r \right) = \left( {1 - \frac{r_h }{r}} \right)\left(
{1 - \frac{r_ - }{r}} \right)^{\frac{1 - a^2}{1 + a^2}}$, $R\left( r
\right) = r\left( {1 - \frac{r_ - }{r}} \right)^{\frac{a^2}{1 +
a^2}}$, $\Phi $ and $F$ are the dilaton and Maxwell fields,
the mass and electric charge of the black hole are expressed as
$M = \frac{r_h }{2} + \frac{r_ - }{2} \cdot \frac{1 - a^2}{1 + a^2}$
and $Q^2 = \frac{r_h r_ - }{1 + a^2}$, respectively. $r_h /r_ - $
are the outer/inner horizon of the black hole, $a$ is a couple
constant confined in $0 \le a < 1$. When
$a = 0$, this metric reduces to the Reissner-Nordstr\"{o}m solution.
The electric potential of the G. H. dilatonic black hole is $A_\mu =
A_t dt = \frac{Q}{r}dt$. For all $a$, it is singular at the location
of the outer horizon $r = r_h $.

The motion equation of charged fermion in the
electromagnetic field can be written as
\begin{equation}
i\gamma ^\mu \left( {\partial _\mu + \Omega _\mu + \frac{i}{\hbar
}eA_\mu } \right)\psi + \frac{m}{\hbar }\psi = 0, \label{eq9}
\end{equation}
where $\Omega _\mu = \frac{i}{2}\Gamma _\mu ^{\alpha \beta } \sum
_{\alpha \beta } $ ,  $\sum _{\alpha \beta } = \frac{i}{4}\left[
{\gamma ^\alpha ,\gamma ^\beta } \right]$ and $\gamma ^\mu $
matrices satisfy $\left\{ \gamma ^\mu ,\gamma ^\nu \right\} =
2g^{\mu \nu }\times I$, $m$ and $e$ are the mass and the electric
charge of the emitted particles, and $A_\mu $ is the electric
potential of the black hole. To deal with fermions tunnelling radiation, it is important to choose an appropriate  $\gamma ^\mu
$ matrices. There are many ways to choose them\cite{M5,
M6}, in our case, we choose \bea &&\gamma ^t = \frac{1}{\sqrt
{e^{2U}\left( r \right)} }\left( {{\begin{array}{*{20}c}
 i \hfill & 0 \hfill \\
 0 \hfill & { - i} \hfill \\
\end{array} }} \right),
\quad \gamma ^\theta = \frac{1}{R\left( r \right)}\left(
{{\begin{array}{*{20}c}
 0 \hfill & {\sigma ^1} \hfill \\
 {\sigma ^1} \hfill & 0 \hfill \\
\end{array} }} \right), \nn\\
&&\gamma ^r = \sqrt {e^{2U}\left( r \right)} \left(
{{\begin{array}{*{20}c}
 0 \hfill & {\sigma ^3} \hfill \\
 {\sigma ^3} \hfill & 0 \hfill \\
\end{array} }} \right),
\quad \gamma ^\phi = \frac{1}{R\left( r \right)\sin \theta }\left(
{{\begin{array}{*{20}c}
 0 \hfill & {\sigma ^2} \hfill \\
 {\sigma ^2} \hfill & 0 \hfill \\
\end{array} }} \right).
\eea Here, $\sigma ^i$ is the Pauli Sigma matrices. For fermion with spin 1/2, the wave function has two spin states
(namely, spin up($\uparrow$) and down ($\downarrow$) states), so we can take the following ansatz as \bea
&&\Psi _ \uparrow = \left( {{\begin{array}{*{20}c}
 A \left({t,r,\theta ,\varphi } \right) \hfill \\
 0 \hfill \\
 B\left({t,r,\theta ,\varphi } \right) \hfill \\
 0 \hfill \\
\end{array} }} \right)\exp \left( {\frac{i}{\hbar }I_ \uparrow \left(
{t,r,\theta ,\varphi } \right)} \right), \nn\\
&& \Psi _ \downarrow = \left( {{\begin{array}{*{20}c}
 0 \hfill \\
 C\left({t,r,\theta ,\varphi } \right) \hfill \\
 0 \hfill \\
 D\left({t,r,\theta ,\varphi } \right) \hfill \\
\end{array} }} \right)\exp \left( {\frac{i}{\hbar }I_ \downarrow \left(
{t,r,\theta ,\varphi } \right)} \right), \label{eq1} \eea where
$\Psi _ \uparrow$ denotes the wave function of spin up particle, and
$\Psi _ \downarrow$ is for spin down case. Inserting Eq.(\ref{eq1})
for spin up particle into the Dirac equation and dividing the
exponential term and multiplying by $\hbar$, we have
\begin{equation}
- \left( {\frac{iA}{\sqrt {e^{2U}\left( r \right)} }\left( {\partial
_t I_ \uparrow + eA_t } \right) + B\sqrt {e^{2U}\left( r \right)}
\partial _r I_ \uparrow } \right) + mA = 0, \label{eqA}
\end{equation}
\begin{equation}
\left( {\frac{iB}{\sqrt {e^{2U}\left( r \right)} }\left( {\partial
_t I_ \uparrow + eA_t } \right) - A\sqrt {e^{2U}\left( r \right)}
\partial _r I_ \uparrow } \right) + mB = 0, \label{eqA1}
\end{equation}
\begin{equation}
{\frac{B}{R\left( r \right)}\partial _\theta I_ \uparrow +
\frac{iB}{R\left( r \right)\sin \theta }\partial _\varphi I_
\uparrow }  = 0,
\end{equation}
\begin{equation}
{\frac{A}{R\left( r \right)}\partial _\theta I_ \uparrow +
\frac{iA}{R\left( r \right)\sin \theta }\partial _\varphi I_
\uparrow }  = 0.
\end{equation}
It is difficult to directly solve the action from above equations.
Considering the symmetries of the space-time, we can
carry out separation of variables for the action as
\begin{equation}
I_\uparrow= - \omega t + W\left( r \right) + \Theta \left( {\theta
,\varphi } \right), \label{eqI}
\end{equation}
where $\omega$ is the energy of the emitted particle. Then
substituting Eq.(\ref{eqI}) into Eqs.(\ref{eqA})and (\ref{eqA1}),
the radial function $W\left( r \right)$ satisfies the following
equations
\begin{equation}
\left( {\frac{iA}{\sqrt {e^{2U}\left( r \right)} }\left( {\omega -
eA_t } \right) - B\sqrt {e^{2U}\left( r \right)}
\partial _r W\left( r \right)} \right) + mA = 0, \label{eq2}
\end{equation}
\begin{equation}
- \left( {\frac{iB}{\sqrt {e^{2U}\left( r \right)} }\left( {\omega -
eA_t } \right) + A\sqrt {e^{2U}\left( r \right)} \partial _r W\left(
r \right)} \right) + mB = 0.\label{eq3}
\end{equation}
Here, we neglect the equations about $\Theta \left({\theta ,\varphi
} \right)$ function. Although $\Theta \left({\theta ,\varphi }
\right)$ could provide a
contribution to the imaginary part of the action, its total
contributions to the tunnelling rate are cancelled out \cite{M5}.
At the event horizon, the radial function $W\left( r \right)$ can be written as
\bea W_\pm \left( r \right)& =& \pm \int {\frac{\sqrt {\left(
{\omega - eA_t } \right)^2 + m^2e^{2U}\left( r \right)}
}{e^{2U}\left( r \right)}dr} \nn\\
&=&\pm i\pi r_h \left( {\omega - eA_0 } \right)\left( {1 - \frac{r_
- }{r_h }} \right)^{\frac{a^2 - 1}{a^2 + 1}}, \label{eqW1} \eea
where $+ $/$-$ correspond to the outgoing/ingoing solutions, and
$A_h  = {Q}/{r_h }$ is the electric potential at the event horizon.
So the tunnelling probability of charged fermion is \bea \Gamma &=&
\frac{P_{\left( {emission} \right)} }{P_{\left( {absorption}
\right)} } = \frac{\exp ( - 2\textrm{Im}I_ {\uparrow+ })}{\exp
( - 2\textrm{Im}I_{\uparrow - })}= \frac{\exp ( - 2\textrm{Im}W_ + )}{\exp ( - 2\textrm{Im}W_ - )}\nn\\
& =& \exp \left( { - 4\pi r_h \left( {\omega - eA_0 } \right)\left(
{1 - \frac{r_ - }{r_h }} \right)^{\frac{a^2 - 1}{a^2 + 1}}} \right),
\eea
which means Hawking temperature of the dilatonic black hole
is
\begin{equation}
T = \frac{1}{4\pi r_h }\left( {1 - \frac{r_ - }{r_h }}
\right)^{\frac{1 - a^2}{1 + a^2}}. \label{T}
\end{equation}

Now Hawking temperature has been correctly derived via fermion with spin up
tunnelling from the dilatonic black hole. For spin down
case, the similar process is adopted and the same result can be
recovered. When $a = 0$, the metric (\ref{eqds}) is the solution of
the Reissner-Nordstr\"{o}m black hole, and Hawking temperature
is recovered from Eq.(\ref{T}) as
\begin{equation}
T = \frac{1}{2\pi }\frac{\sqrt {M^2 - Q^2} }{\left( {M + \sqrt {M^2
- Q^2} } \right)^2}. \label{eq17}
\end{equation}
In the extreme limit $r_h = r_ - $, it is found that the surface
gravity is zero and Hawking temperature of the black holes (the
extreme Reissner-Nordstr\"{o}m black hole and the extreme $U\left( 1
\right)$ dilatonic black hole) vanishes for all $0 \le a < 1$. But
for $a = 1$ the surface gravity is a constant; and for $a > 1$, it
divergences because the black hole approaches its extremal
limit\cite{M8}, at the time the nearly extremal black hole behaves
more like elementary particles.
%%%%%%%%%%%%%%%%%%%%%%%%%%%%%%%%%%%%%%%%%%%%%%%%%%%%%%%%%%%%%
\section{Fermions tunnelling from the rotating EMDA black hole}\label{EMDA}
%%%%%%%%%%%%%%%%%%%%%%%%%%%%%%%%%%%%%%%%%%%%%%%%%%%%%%%%%%%%%
In this section, we study charged fermions tunnelling from the
rotating Einstein-Maxwell-Dilaton-Axion(EMDA) black hole. In 1995,
Alberto Garc\'{\i}a et.al gave a class of stationary axisymmetric
solutions of the Einstein-Maxwell-Dilaton-Axion field equations.
From the action \bea S &=& \int dx^4\sqrt { - g} [ R - 2g^{\mu
\nu}\partial _\mu \phi
\partial _v \phi \nn\\
&-& \frac{1}{2}e^{4\phi }g^{\mu \nu }\partial _\mu \kappa
\partial _v \kappa - e^{ - 2\phi }F_{\mu \nu } F^{\mu \nu } - \kappa F_{\mu
\nu }
\mathord{\buildrel{\lower3pt\hbox{$\scriptscriptstyle\smile$}}\over
{F}} ^{\mu \nu } ] , \eea the solution of the EMDA black
hole\cite{M9} can be obtained as
\bea ds^2 &=& - \frac{\Delta -
a^2\sin ^2\theta }{\sum }dt^2 - \frac{2a\sin ^2\theta \left( {r^2
+ 2br + a^2 - \Delta } \right)}{\sum }dtd\varphi \nn\\
&+& \frac{\sum}{\Delta }dr^2 + \sum d\theta ^2 +\frac{\left( {r^2 +
2br + a^2} \right)^2 - \Delta a^2\sin ^2\theta }{\sum }\sin ^2\theta
d\varphi ^2, \label{eq16} \eea where \bea
&& A_\mu =A_t dt + A_\varphi d\varphi = \frac{Qr}{\sum}dt - \frac{Qra\sin ^2\theta }{\sum }d\varphi ,\nn\\
&& \sum = r^2 + 2br + a^2\cos ^2\theta, \nn\\
&& \Delta = r^2 - 2mr + a^2 = \left( {r - r_h } \right)\left( {r - r_ - }\right),\nn\\
&& r_h = m + \sqrt{m^2 - a^2}, r_ - = m - \sqrt{m^2- a^2}. \eea
The dilaton $\phi $ and axion scalar $\kappa $ fields are,
respectively, given as $\exp \left( {2\phi _0 } \right) = \omega $
and $\kappa = \kappa _0 $, where $\omega $ and $\kappa _0 $ are
constants, and $r_h $/$r_ -$  are the outer/inner horizons of
the black hole. The parameters $m$, $a$ and $b$ are the mass,
angular momentum per unit mass and dilatonic constant of the black
hole, which are related to the ADM mass $M$, charge $Q$ and
angular momentum $J$ of the black hole with
\begin{equation}
M = m + b, ~~Q^2 = 2b\left( {m + b} \right),~~J =\left( {m + b}
\right)a.
\end{equation}
When $a = 0$, the EMDA metric reduces to the
Garfinkle-Horowitz-Strominger dilatonic solution. When $b = m\sinh
^2\left(\frac{\alpha}{2}\right)$ and $\omega = 1$, one can derive
the parameters of characterizing the Kerr-Sen black hole. For
simplicity of our computation, we introduce the dragging coordinate
transformation as $\phi = \varphi -\Omega t$, where
\begin{equation}
\Omega = \frac{a\left( {r^2 + 2br + a^2 - \Delta } \right)}{\left(
{r^2 + 2br + a^2} \right)^2 - \Delta a^2\sin ^2\theta},
\end{equation}
to the metric (\ref{eq16}), then the new metric takes the forms as
\begin{equation}
ds^2 = - f\left( r \right)dt^2 + \frac{1}{g\left( r \right)}dr^2 +
\sum d\theta ^2 + g_{33} d\phi ^2, \label{eq20}
\end{equation}
where \bea && f\left( r \right) = \frac{\Delta \sum }{\left( {r^2 +
2br + a^2}
\right)^2 - \Delta a^2\sin ^2\theta }, \quad g\left( r\right) = \frac{\Delta }{\sum }, \nn\\
&& g_{33} = \frac{\left( {r^2 + 2br + a^2} \right)^2 - \Delta
a^2\sin ^2\theta }{\sum }\sin ^2\theta . \eea
The corresponding potential is
\begin{equation}
\mathcal{A}_\mu = \mathcal{A}_t dt + \mathcal{A}_\phi d\phi =
\frac{\left( {r^2 + 2br + a^2} \right)Qr}{\left( {r^2 + 2br + a^2}
\right)^2 - \Delta a^2\sin ^2\theta }dt - \frac{Qra\sin ^2\theta
}{\sum }d\phi .
\end{equation}
In order to solve Dirac equation, we must first introduce the
metrics $\gamma ^\mu $. As mentioned in Sec.\ref{SSDB}, there are
many different ways to choose them. Considering the similarity between
the metrics (\ref{eqds}) and (\ref{eq20}), we choose $\gamma ^\mu $
matrices as
\bea && \gamma ^t = \frac{1}{\sqrt {f\left( r \right)} }\left(
{{\begin{array}{*{20}c}
 i \hfill & 0 \hfill \\
 0 \hfill & { - i} \hfill \\
\end{array} }} \right),
\quad \gamma ^\phi = \frac{1}{\sqrt {g_{33} } }\left(
{{\begin{array}{*{20}c}
 0 \hfill & {\sigma ^2} \hfill \\
 {\sigma ^2} \hfill & 0 \hfill \\
\end{array} }} \right), \nn\\
&& \gamma ^r = \sqrt {g\left( r \right)} \left(
{{\begin{array}{*{20}c}
 0 \hfill & {\sigma ^3} \hfill \\
 {\sigma ^3} \hfill & 0 \hfill \\
\end{array} }} \right),
\quad \gamma ^\theta = \frac{1}{\sqrt {\sum }}\left(
{{\begin{array}{*{20}c}
 0 \hfill & {\sigma ^1} \hfill \\
 {\sigma ^1} \hfill & 0 \hfill \\
\end{array} }} \right). \label{eq23}
\eea
We still choose the same form as given in Eq.(\ref{eq1}) for the
wave functions of charged fermion tunnelling from the EMDA black
hole, and only explore spin up case. Substituting the $\gamma ^\mu $
matrices (\ref{eq23}) and the wave function into Dirac equation
(\ref{eq9}), we have
\begin{equation}
 - \left( {\frac{iA}{\sqrt {f\left( r \right)} }\left( {\partial _t I_
\uparrow + e\mathcal{A}_t } \right) + B\sqrt {g\left( r \right)}
\partial _r I_ \uparrow } \right) + mA = 0,\label{eq27}
\end{equation}
\begin{equation}
\left( {\frac{iB}{\sqrt {f\left( r \right)} }\left( {\partial _t I_
\uparrow + e\mathcal{A}_t } \right) - A\sqrt {g\left( r \right)}
\partial _r I_ \uparrow } \right) + mB = 0, \label{eq28}
\end{equation}
\begin{equation}
 {\frac{B}{\sqrt {\sum }}\partial _\theta I_ \uparrow +
\frac{iB}{\sqrt {g_{33} } }\left( {\partial _\phi I_ \uparrow +
e\mathcal{A}_\phi } \right)} = 0,
\end{equation}
\begin{equation}
{\frac{A}{\sqrt {\sum }}\partial _\theta I_ \uparrow +
\frac{iA}{\sqrt {g_{33} } }\left( {\partial _\phi I_ \uparrow +
e\mathcal{A}_\phi } \right)} = 0.
\end{equation}
There are four equations, but our interest is the first two ones
because the tunnelling rate is directly related to the imaginary
part of the radial function, and the angular contribution can be
cancelled out when dividing the outgoing probability by the ingoing
probability. In view of the properties of the rotating EMDA
space-time, one can carry out separation of variables for the action as
\begin{equation}
I_ \uparrow = - \left(\omega-j\Omega\right) t + W\left( r \right) +
j\phi + \Theta \left( \theta \right), \label{eq31}
\end{equation}
where $\omega $ is the energy of the emitted particle for the
observer at the infinity, and $j$ is the angular quantum number
about $\varphi$. Inserting the action (\ref{eq31}) into Eqs.
(\ref{eq27}) and (\ref{eq28}) yields
\begin{equation}
\left( {\frac{iA}{\sqrt {f\left( r \right)} }\left( {\omega -
j\Omega - e\mathcal{A}_t } \right) - B\sqrt {g\left( r \right)}
\partial _r W\left( r \right)} \right) + mA = 0,
\end{equation}
\begin{equation}
 - \left( {\frac{iB}{\sqrt {f\left( r \right)} }\left( {\omega - j\Omega -
e\mathcal{A}_t } \right) + A\sqrt {g\left( r \right)} \partial _r
W\left( r \right)} \right) + mB = 0.
\end{equation}
Here exists two cases. When $m=0$, the above equations describe the
radial wave function for the massless particle. When $m \ne 0$,
charged massive fermion is in consideration, and solving the above
equations yields
\bea W_\pm \left( r \right) &=& \pm \int {\sqrt
{\frac{\left( {\omega - j\Omega - e\mathcal{A}_t } \right)^2+ m^2f
\left( r
\right)}{f\left( r \right)g\left( r \right)}} dr} \nn\\
& =&  \pm i\pi\frac{\omega - j\Omega _h - e\mathcal{A}_t(r_h)
}{\sqrt {{f}'\left( {r_h } \right){g}'\left( {r_h } \right)} }, \eea
where $+ $($-$) correspond to the outgoing (ingoing) solutions, and
$\Omega _h=\Omega(r_h)=a/\left(r_h^2+2br_h+a^2\right)$ is the
angular velocity at the event horizon of the EMDA black hole. Thus
the tunnelling probability of charged fermion can be written as \bea
\Gamma &=& \frac{P_{\left( {emission} \right)} }{P_{\left(
{absorption} \right)} } = \frac{\exp ( - 2\textrm{Im}I_ {\uparrow+}
)}{\exp
( - 2\textrm{Im}I_ {\uparrow- })}=\frac{\exp ( - 2\textrm{Im}W_ + )}{\exp( - 2\textrm{Im}W_ - )} \nn\\
&=& \exp \left( { - 4\pi \frac{\omega - j\Omega _h -
e\mathcal{A}_t(r_h) }{\sqrt {{f}'\left( {r_h } \right){g}'\left(
{r_h } \right)} }} \right). \eea According to the relationship
between the tunnelling rate and Hawking temperature,  Hawking
temperature of the EMDA black hole can be obtained as
\begin{equation}
T = \frac{\sqrt {{f}'\left( {r_h } \right){g}'\left( {r_h } \right)}
}{4\pi } = \frac{1}{2\pi }\frac{r_h - m}{r_h^2 + 2br_h + a^2}.
\label{eq37}
\end{equation}

When $b = 0$, Hawking temperature $T$ can be written as
\begin{equation}
T
=\frac{1}{2\pi}\frac{\sqrt{m^2-a^2}}{\left(m+\sqrt{m^2-a^2}\right)^2+a^2},
\label{eq38}
\end{equation}
which is Hawking temperature of the Kerr black hole. For $b = 0$
and $a = 0$ (Schwarzschild black hole case), Hawking temperature equals $1/8\pi m $. So fermions tunnelling from
these black hole can correctly recover Hawking temperatures.

%%%%%%%%%%%%%%%%%%%%%%%%%%%%%%%%%%%%%%%%%%%%%%%%%%%%%%%%%%%%%
\section{Fermions tunnelling from the rotating Kaluza-Klein black hole}\label{KK}
%%%%%%%%%%%%%%%%%%%%%%%%%%%%%%%%%%%%%%%%%%%%%%%%%%%%%%%%%%%%%

The solution of the rotating Kaluza-Klein black hole \cite{M10}
denoted a five dimensional space-time with a translational
symmetry in a spacelike direction in four dimensional metrics
$g_{\mu \nu } $ is obtained from the action (\ref{S2}) with $a =
\sqrt 3 $, which reads \bea ds^2 &=& - \frac{\Delta - a^2\sin
^2\theta }{\Pi \sum }dt^2 +
\frac{\Pi \sum }{\Delta }dr^2 + \Pi \sum d\theta ^2 \nn\\
&+& \left[ {\Pi \left( {r^2 + a^2} \right) +\frac{{\rm Z}}{\Pi
}a^2\sin ^2\theta } \right]\sin ^2\theta d\varphi ^2 - \frac{2a{\rm
Z}\sin ^2\theta }{\Pi \sqrt {1 - v^2} }dtd\varphi , \eea with the
dilaton field $\Phi =- (\sqrt{3}\ln \Pi)/2$, and the electromagnetic
potential
\begin{equation}
A_\mu = \frac{v}{2\left( {1 - v^2} \right)}\frac{{\rm Z}}{\Pi ^2}dt
- \frac{va\sin ^2\theta }{2\sqrt {1 - v^2} }\frac{{\rm Z}}{\Pi
^2}d\varphi ,
\end{equation}
where \bea && Z = \frac{2mr}{\sum }, \quad \Pi = \sqrt {1 +
\frac{v^2{\rm Z}}{1 - v^2}} , \nn\\
&& \sum = r^2 + a^2\cos ^2\theta ,
\quad \Delta = r^2 - 2mr + a^2, \nn\\
&& r_h = m + \sqrt {m^2 - a^2}, \quad r_ - = m - \sqrt {m^2 - a^2}.
\eea and $a$ and $v$ are the rotation parameter and  boost
velocity respectively. The outer(inner) horizons are described by
the parameters $r_h$($r_ -$), and satisfy $\Delta = 0$. The solution
reduces to the Kerr solution for $v=0$. The physics mass $M$, charge
$Q$ and angular momentum $J$ of the black hole are related to the
parameters $m$, $a$ and $v$ as
\begin{equation}
M = \frac{m}{2} \cdot \frac{2 - v^2}{1 - v^2},~~ Q = \frac{mv}{1 -
v^2},~~J = \frac{ma}{\sqrt {1 - v^2} }.
\end{equation}
As mentioned in Sec.\ref{EMDA}, to easily investigate charged fermions tunnelling from the black hole, we first
introduce the dragging coordinate transformation as $\phi = \varphi
- \Omega t$, where
\begin{equation}
\Omega = \frac{a{\rm Z}}{\left[ {\Pi ^2\left( {r^2 + a^2} \right) +
{\rm Z}a^2\sin ^2\theta } \right]\sqrt {1 - v^2} }.
\end{equation}
Then the new metric takes the form as
\begin{equation}
ds^2 = - F\left( r \right)dt^2 + \frac{1}{G\left( r \right)}dr^2 +
\Pi \sum d\theta ^2 + g_{33} d\phi ^2, \label{eq43}
\end{equation}
with the new electromagnetic potential
\bea
\mathcal{A}_\mu &=& \mathcal{A}_t dt + \mathcal{A}_\phi d\phi \nn\\
&=& \frac{Qr}{\sum }\frac{r^2 + a^2}{\Pi ^2\left( {r^2 + a^2}
\right)^2 + {\rm Z}a^2\sin ^2\theta }dt - \frac{va\sin ^2\theta
}{2\sqrt {1 - v^2} }\frac{{\rm Z}}{\Pi ^2}d\phi , \eea and \bea
&&F\left( r \right) = - \frac{\Pi \sum \Delta \left( {1 - v^2}
\right)}{\left( {r^2 + a^2} \right)^2 - \Delta \left( {a^2\sin
^2\theta + v^2\sum } \right)}, \nn\\
&&g_{33} = \left( {\Pi \left( {r^2 + a^2} \right) + \frac{{\rm
Z}}{\Pi }a^2\sin ^2\theta } \right)\sin ^2\theta , \quad G\left( r
\right) = \frac{\Delta }{\Pi \sum }. \eea
As the metric (\ref{eq43}) takes the similar form as that of (\ref{eq20}),
we can choose the similar $\gamma ^\mu $ matrices as Eq.(\ref{eq23}) for the black hole,
specifically takeing
\bea &&\gamma ^t = \frac{1}{\sqrt {F\left( r
\right)} }\left( {{\begin{array}{*{20}c}
 i \hfill & 0 \hfill \\
 0 \hfill & { - i} \hfill \\
\end{array} }} \right),
\quad \gamma ^\theta = \frac{1}{\sqrt {\Pi \sum } }\left(
{{\begin{array}{*{20}c}
 0 \hfill & {\sigma ^1} \hfill \\
 {\sigma ^1} \hfill & 0 \hfill \\
\end{array} }} \right), \nn\\
&&\gamma ^r = \sqrt {G\left( r \right)} \left(
{{\begin{array}{*{20}c}
 0 \hfill & {\sigma ^3} \hfill \\
 {\sigma ^3} \hfill & 0 \hfill \\
\end{array} }} \right),
\quad \gamma ^\phi = \frac{1}{\sqrt {g_{33} } }\left(
{{\begin{array}{*{20}c}
 0 \hfill & {\sigma ^2} \hfill \\
 {\sigma ^2} \hfill & 0 \hfill \\
\end{array} }} \right).
\eea

Substituting the above $\gamma ^\mu $ matrices and the wave function
with spin up given in (\ref{eq1}) into Dirac equation yields
\begin{equation}
 - \left( {\frac{iA}{\sqrt {F\left( r \right)} }\left( {\partial _t I_
\uparrow + e\mathcal{A}_t } \right) + B\sqrt {G\left( r \right)}
\partial _r I_ \uparrow } \right) + mA = 0,\label{eq47}
\end{equation}
\begin{equation}
\left( {\frac{iB}{\sqrt {F\left( r \right)} }\left( {\partial _t I_
\uparrow + e\mathcal{A}_t } \right) - A\sqrt {G\left( r \right)}
\partial _r I_ \uparrow } \right) + mB = 0,\label{eq48}
\end{equation}
\begin{equation}
{\frac{B}{\sqrt {\Pi \sum } }\partial _\theta I_ \uparrow +
\frac{iB}{\sqrt {g_{33} } }\left( {\partial _\phi I_ \uparrow +
e\mathcal{A}_\phi } \right)}  = 0,
\end{equation}
\begin{equation}
{\frac{A}{\sqrt {\Pi \sum } }\partial _\theta I_ \uparrow +
\frac{iA}{\sqrt {g_{33} } }\left( {\partial _\phi I_ \uparrow +
e\mathcal{A}_\phi } \right)}  = 0.
\end{equation}
Although there are four equations, our attention is also focused on
the first two equations. Considering the properties of the
Kaluza-Klein space-time, we carry out separation of variables as
Eq.(\ref{eq31}). Inserting the action $I_ \uparrow $ into
Eqs.(\ref{eq47}) and (\ref{eq48}) yields
\begin{equation}
 \left( {\frac{iA}{\sqrt {F\left( r \right)} }\left( {\omega - j\Omega -
e\mathcal{A}_t } \right) - B\sqrt {G\left( r \right)} \partial _r
W\left( r \right)} \right) + mA = 0,
\end{equation}
\begin{equation}
 - \left( {\frac{iB}{\sqrt {F\left( r \right)} }\left( {\omega - j\Omega -
e\mathcal{A}_t } \right) + A\sqrt {G\left( r \right)} \partial _r
W\left( r \right)} \right) + mB = 0,
\end{equation}
where $\omega$ denotes the energy of the emitted particles measured
by the observer at the infinity, and $j$ is the angular quantum
number about $\varphi$. In the case $m \ne 0$, solving the above
equations, we have
\bea W_\pm \left( r
\right) &=& \pm \int {\sqrt {\frac{\left( {\omega - j\Omega -
e\mathcal{A}_t } \right)^2 +m^2F\left( r
\right)}{F\left( r \right)G\left( r \right)}} dr} \nn\\
&=& \pm i\pi \frac{\omega - j\Omega _h - e\mathcal{A}_t(r_h) }{\sqrt
{{F}'\left( {r_h } \right){G}'\left( {r_h } \right)} }, \eea where
$+(-) $ signs represent the outgoing(ingoing) solutions, and $\Omega
_h =\Omega(r_h)$ is the angular velocity at the outer horizon of the
KK black hole. So the tunnelling probability of charged fermion of
the black hole can be written as \bea \Gamma &=& \frac{P_{\left(
{emission} \right)} }{P_{\left( {absorption} \right)} } = \frac{\exp
( - 2\textrm{Im}I_ {\uparrow+} )}{\exp
( - 2\textrm{Im}I_ {\uparrow- })}=\frac{\exp ( - 2\textrm{Im}W_ + )}{\exp( - 2\textrm{I}mW_ - )} \nn\\
&=& \exp \left( { - 4\pi \frac{\omega - j\Omega _h -
e\mathcal{A}_t(r_h) }{\sqrt {{F}'\left( {r_h } \right){G}'\left(
{r_h } \right)} }} \right). \eea

Thus Hawking temperature of the Kaluza-Klein black hole takes
the form as
\begin{equation}
 T = \frac{\sqrt {{F}'\left( {r_h } \right){G}'\left(
{r_h } \right)} }{4\pi } = \frac{1}{4\pi } \cdot \frac{\sqrt {\left(
{1 - v^2} \right)\left( {m^2 - a^2} \right)} }{m\left( {m + \sqrt
{m^2 - a^2} } \right)}. \label{eq56}
\end{equation}

When $v = 0$ and $Q=0$, Hawking temperature of Eq.(\ref{eq56})
equals Eq.(\ref{eq38}), which means, in that case, the KK black hole
can reduce to the Kerr black hole. And for $v = 0$ and $a = 0$, it
describes a Schwarzschild black hole, and its Hawking temperature
equals $1/8\pi m$. Obviously, the results once again proves the validity of the charged
fermions tunnelling method.

%%%%%%%%%%%%%%%%%%%%%%%%%%%%%%%%%%%%%%%%%%%%%%%%%%%%%%%%%%%%%
\section{Conclusions and Discussions}\label{cd}
%%%%%%%%%%%%%%%%%%%%%%%%%%%%%%%%%%%%%%%%%%%%%%%%%%%%%%%%%%%%%

In this paper, we attempted to apply Kerner and Man's fermions
tunnelling method to charged fermions' cases. As an example, Hawking
radiation of charged fermions for a general charged, spherically
symmetric dilatonic black hole is first studied via the tunnelling
method. For a wide extension, charged fermions tunnelling from the
rotating dilatonic black holes, specifically including the rotating
Einstein-Maxwell-Dilaton-Axion (EDMA) and  Kaluza-Klein (KK) black
holes, are also considered in the paper. As a result, the correct
Hawking temperatures are well described by charged fermions
tunnelling from these black holes.

For simplicity to choose the metrics $\gamma^\mu$ when dealing
with Hawking radiation of charged fermionss tunnelling from the
rotating dilatonic black holes in Sec.(\ref{EMDA}) and
Sec.(\ref{KK}), we carried out the dragging coordinate
transformation. In fact, such behavior would not
stop us from getting the correct Hawking temperature of these black
holes, as in Ref.\cite{M5} to discuss Hawking radiation of black holes in the Painlev\'{e} and
Kruskal-Szekers coordinate systems. In addition, after charged
fermionss have tunnelled out, we assumed that the energy, charge and
angular momentum of the dilatonic black holes keep the same as
before. If the emitted particle's self-gravitational interaction is
incorporated into Hawking radiation of fermionss tunnelling, Hawking
temperatures will be corrected slightly, but their leading
terms take the same form as Eqs.(\ref{eq17}),(\ref{eq37}) and
(\ref{eq56}).

In summary, we have succeeded in dealing with Hawking radiation of
the rotating black holes via charged fermions tunnelling. This method
can also be directly extended to the case of the non-stationary
charged rotating black holes.

\section*{Acknowledgments}
This work was partially supported by the Natural Science Found of
China under Grant Nos.10675051, 10705008 and 10773008, and a
Graduate Innovation Foundation by CCNU.

\end{document}